\documentclass[12pt,preprint]{aastex}
\def\p/{\mbox{$^1$}}
\def\pp/{\mbox{$^2$}}
\def\ppp/{\mbox{$^3$}}
\def\pppp/{\mbox{$^4$}}
\def\m/{\mbox{$^{-1}$}}
\def\mm/{\mbox{$^{-2}$}}
\def\mmm/{\mbox{$^{-3}$}}
\def\mmmm/{\mbox{$^{-4}$}}
\def\Ms/{\mbox{M$_\odot$}}

\def\bt{\mbox{$B_T$}}
\def\vt{\mbox{$V_T$}}
\def\btvt{\mbox{$B_T-V_T$}}

\slugcomment{Accepted at Publications of the Astronomical Society of the Pacific (June 2005 issue)}
\begin{document}

\title{A cross-calibration between Tycho-2 photometry and HST spectrophotometry}
\shorttitle{A cross-calibration between Tycho-2 photometry and HST spectrophotometry}

\author{J. Ma\'{\i}z Apell\'aniz\altaffilmark{1}}
\affil{Space Telescope Science Institute\altaffilmark{2}, 3700 San Martin 
Drive, Baltimore, MD 21218, U.S.A.}
\email{jmaiz@stsci.edu}


\altaffiltext{1}{Affiliated with the Space Telescope Division of the European 
Space Agency, ESTEC, Noordwijk, Netherlands.}
\altaffiltext{2}{The Space Telescope Science Institute is operated by the
Association of Universities for Research in Astronomy, Inc. under NASA
contract No. NAS5-26555.}

\begin{abstract}
I show that Tycho-2 photometry and HST spectrophotometry are accurate and stable enough to
obtain a precise cross-calibration by analyzing a well-calibrated sample of 256 stars observed 
with both Hipparcos and HST. 
Based on this analysis, I obtain the following photometric zero points in the Vega 
magnitude system for Tycho-2: 0.020$\pm$0.001 (\btvt), 0.078$\pm$0.009 (\bt), and
0.058$\pm$0.009 (\vt).
\end{abstract}


\section{Introduction}

	It is nowadays extremely common to analyze photometric observations by comparing them
with synthetic photometry derived from observed or theoretical spectral energy distributions
(SEDs).  Such comparisons are done by establishing a magnitude system based on a reference 
spectrum, be that of a real object, such as Vega, or a simple artificial spectrum, such as a 
constant in $f_\lambda$ or $f_\nu$ \citep{synphot}. Each reference spectrum yields a different 
magnitude system defined from the expression:

\begin{equation}
m_{P} = -2.5\log_{10}\left(\frac{\int P(\lambda)f_{\lambda}(\lambda)\lambda\,d\lambda}
                                {\int P(\lambda)f_{\lambda{\rm,ref}}(\lambda)\lambda\,d\lambda}\right)
                                + {\rm ZP}_P,
\end{equation}

\noindent where $P(\lambda)$ is the total-system sensitivity curve, $f_\lambda(\lambda)$ is the
SED of the object, $f_{\lambda{\rm,ref}}(\lambda)$ is the SED of the reference spectrum, and ZP$_P$ 
is the zero point for filter $P$. In order to attain accurate results when comparing measured 
magnitudes or colors with SED models, it is necessary not only to have an accurate knowledge of 
$f_{\lambda{\rm,Vega}}(\lambda)$ but also of $P(\lambda)$ and of ZP$_P$. For example, 
\citet{BohlGill04} measured the spectrum of Vega and found that in the Johnson system (which 
uses that star as a reference), ZP$_V$ = 0.026$\pm$0.008. One can also define a system where
ZP$_P$ = 0 for any filter, such as the VEGAMAG\footnote{Note that it is possible to derive
VEGAMAG magnitudes from SEDs using the Johnson throughputs but those values should be different
from published photometry due to the different zero points used.}, STMAG, or ABMAG systems
\citep{synphot}.

	The lack of an accurate
knowledge of the system sensitivities has plagued some ground-based systems such as Johnson's
$UBV$ \citep{Bessetal98}, mostly due to the use of different observing conditions, technologies,
and reduction techniques by different observers. 
Modern space-based telescopes tend to be more uniform and stable and
usually produce better-quality photometry. ZP$_P$ is usually close to 0.0 but not exactly so,
the reason being the common use of 
secondary calibrators (Vega is often too bright for a detector and/or 
unaccesible from many sites to be observed).
ZP$_P$ thus needs to be measured \citep{Bessetal98,Coheetal03,BohlGill04}, since 
assuming it is exactly zero can introduce systematic errors in the comparison between observed
magnitudes and reference SEDs. 

	The Hipparcos mission \citep{ESA97} observed the full sky and yielded the Tycho 
catalog, which is reasonably complete down to $V = 11.5$. The original Tycho catalog was 
consequently reprocessed by \citet{Hogetal00a} to produce the Tycho-2 catalog, which contains
2.5 million stars and is currently the most complete and accurate all-sky photometric survey in
the optical. The Tycho-2 catalog contains photometry in two optical bands, \bt\ and \vt,
whose sensitivities were analyzed by \citet{Bess00} and use Vega as the reference spectrum. 
The careful processing of Tycho-2 
photometry was described by \cite{Hogetal00b}, including the difffferent tests used to check for
possible systematic errors. 

	HST spectrophotometry is calibrated using the method of \citet{ColiBohl94}, which
is based on a combination of Landolt $BV$ photometry, ground-based spectrophotometry, and SED
models. Four white dwarf stars observed with the Space Telescope Imaging Spectrograph (STIS) 
are used as primary calibrators: G191B2B, GD71, GD153, and HZ43 \citep{Bohletal95}. The absolute 
flux calibration has an accuracy of 4\% in the FUV and 2\% in
the optical \citep{Bohl00} but, given that the photometric repeatibility of STIS is 
0.2-0.4\% \citep{Bohletal01}, the relative flux calibration for colors derived from STIS spectra 
is expected to be better than 2\% in the optical. The latest values for the 
Advanced Camera for Surveys (ACS) photometric zero 
points are also based on that same calibration \citep{DeMaetal04}.

	Given the usefulness of the Tycho-2 photometric database for a number of astronomical 
studies and the accuracy of the absolute flux calibration of HST spectrophotometry, I
considered it important to calibrate the first with respect to the second one, as well as to check
the existence of possible problems in either of them. Ultimately, we want to obtain
the zero points for Tycho-2 magnitudes, ZP$_{B_T}$ and ZP$_{V_T}$, and color, ZP$_{B_T-V_T}$.

\section{The data}

	In order to test the consistency of the Tycho-2 and HST calibrations we would ideally
need a large uniform sample of stars with \bt\ and \vt\ magnitudes that has also been observed
in the optical with an HST spectrograph. Such a sample has indeed been obtained for the Next
Generation Spectral Library (or NGSL, \citealt{Gregetal04}, see also
{\tt http://lifshitz.ucdavis.edu/\~{}mgregg/gregg/ngsl/ngsl.html}). Originally, the NGSL
intended to obtain low-resolution 1660-10\,200 \AA\ spectra for 600 stars with a wide range of 
temperatures, gravities, and metallicities. Unfortunately, the failure of STIS in August 2004
ended the program before reaching that goal. However, 378 stars that are also included in the 
Tycho-2 catalog had already been observed, and that number is large enough for our purposes. 

\begin{figure}
\centerline{\includegraphics*[width=0.47\linewidth]{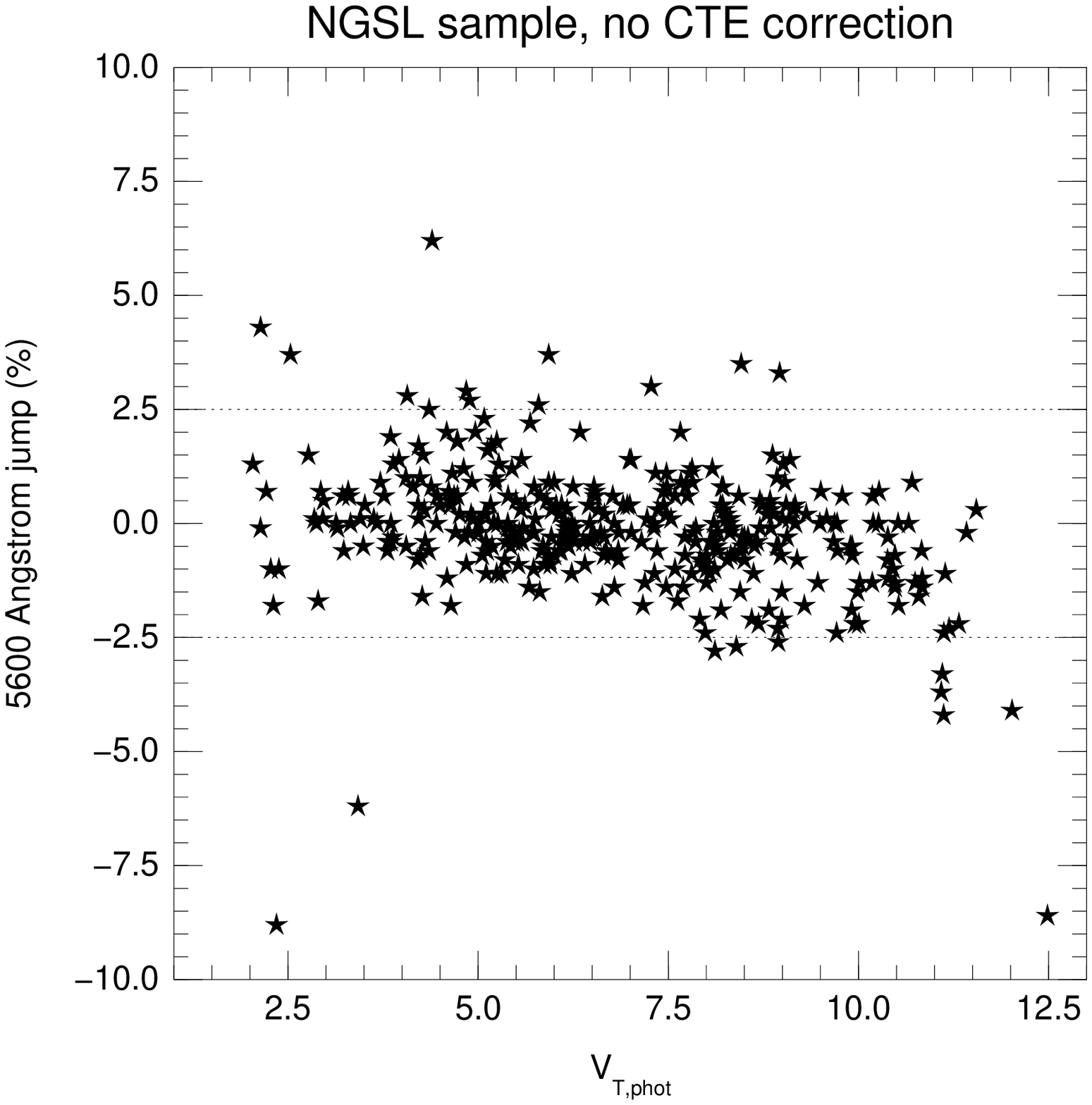}
            \includegraphics*[width=0.47\linewidth]{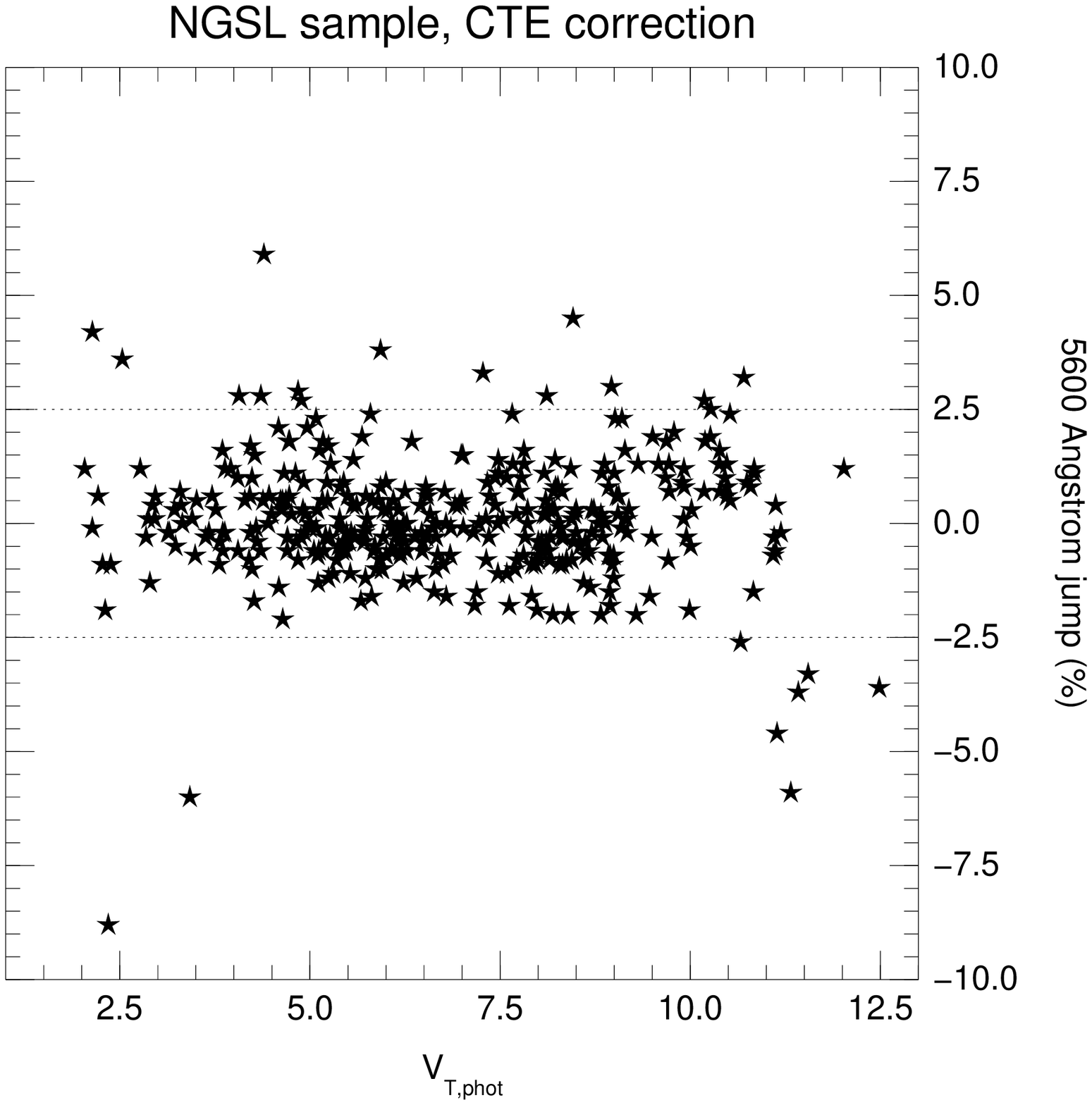}}
\caption{The 5600 \AA\ flux discontinuity as a function of \vt\ (photometric) without (left)
and with (right) CTE correction for the 378 stars in the NGSL sample. A single object is 
located off the scale shown. The dashed lines indicate the limits considering for defining the
sample used to derive ZP$_{B_T-V_T}$}
\label{jump5600}
\end{figure}

	Each star was observed using three different STIS gratings, G230LB, G430L, and G750L.
Here, only the latter two will be used since those are the ones that span the Tycho-2 filters. The
NGSL spectra were obtained using the 52$\times$0.2 slit without a peak-up, which is not the ideal
configuration for absolute spectrophotometric calibration, the reason being that the slit is 
narrow enough that if the star is not well centered, the aperture correction used may not be
accurate. In principle, an off-center positioning could also cause problems with the 
relative spectrophotometric calibration if the aperture correction was strongly dependent on
wavelength due to variations in the PSF\footnote{Also, due to diffraction off the slit edge.}. 
Fortunately, for G430L and G750L that is only a strong effect at the long-wavelength 
part of the spectrum beyond \vt\ \citep{stis}. Therefore, the relative fluxes (and hence the 
derived colors) should not be affected as long as the star did not shift 
position significantly during the visit. Given that all exposures for a given star were obtained 
in a single orbit (usually not filling it completely) and that the typical drift rate for HST 
is less than 10 mas/hr \citep{stis}, indeed one would expect a shift of less than 10\% of a
STIS CCD pixel in the relative position of the star with respect to the slit between the first 
and the last exposure. There are also two calibration issues that need to be addressed regarding
the fact that all exposures were obtained at the E1 position using a CCD subarray. The use of
a subarray requires the definition of a background region closer to the star than for a 
measurement with a full array; I checked that this selection did not affect the measured 
fluxes. Also, the location of the E1 position, closer to the readout point in the detector,
minimizes CTE effects but the CTE correction algorithm itself \citep{BohlGoud03} was not 
originally designed to work with CCD subarrays, so a test should be carried out to check for its
validity.

	In order to explore the possible problems with the spectrophotometric accuracy 
mentioned in the previous paragraph, I first calculated the effect of the CTE correction on the
derived spectrophotometric \bt\ and \vt\ magnitudes by extracting the spectra with and without
the correction and comparing the results. The correction turned out to be very small,
with an average of $-$0.010 magnitudes for \vt, $-$0.018 magnitudes for \bt, and $-$0.008 
magnitudes for \btvt. As a further
check, I measured the value of the discontinuity between the G430L and G750L gratings by 
calculating the ratio between the two fluxes in the 5550-5650 \AA, region, where both gratings 
overlap in coverage. Results are shown in Fig.~\ref{jump5600}, both without applying the CTE 
correction and applying it. As it can be seen, in both cases the majority of the data points are
clustered around a zero value for the jump, indicating the lack of a systematic effect.
However, for the non-CTE-correction case a small slope as a function of \vt\ is present, which
indicates that a non-negligible correction is indeed required. For the case with CTE correction,
the data have no dependency with \vt\ and have a distribution with mean of 0.12\% and standard
deviation of 1.83\%.

	Given the results in the previous paragraph, I adopted the following criteria for
selecting the NGSL sample:

\begin{itemize}
  \item The CTE correction was applied.
  \item Only stars with a 5600 \AA\ jump of less than 2.5\% (in absolute value) were included.
  \item Given that the spectrophotometry and the photometry used in this article were obtained
	in different epochs, I eliminated variable stars. The criterion used was the presence
 	of a variability flag in the Tycho photometry.
\end{itemize}

	With the restrictions above, our final NGSL sample was reduced to 256 objects.

	In order to provide a further check on the possible effects of systematic aperture and
CTE effects I used a control sample by analyzing the spectrophotometric standards of
\citet{Bohletal01} that have Tycho-2 photometry with magnitudes brighter than 12.0 and 
uncertainties of 0.06 magnitudes or less. Those stars were observed with STIS using the 
52$\times$2 slit and with the full CCD array, so the problems mentioned above (slit centering
and validity of the CTE correction algorithm) should not be 
relevant. Four of the \citet{Bohletal01} stars satisfy the above requirements for both 
\bt\ and \vt\ (BD +17 4708, BD +28, 4211, BD +75 325, HD 93521) and an additional three 
do so only for \bt\ (AGK + 81 266, Feige 34, HZ 44).

	The stars in both samples were analyzed using a synthetic photometry code created 
for \citet{Maiz04c} and their spectrophotometric \bt\ and \vt\ magnitudes calculated.

\section{Results}

\begin{figure}
\centerline{\includegraphics*[width=\linewidth]{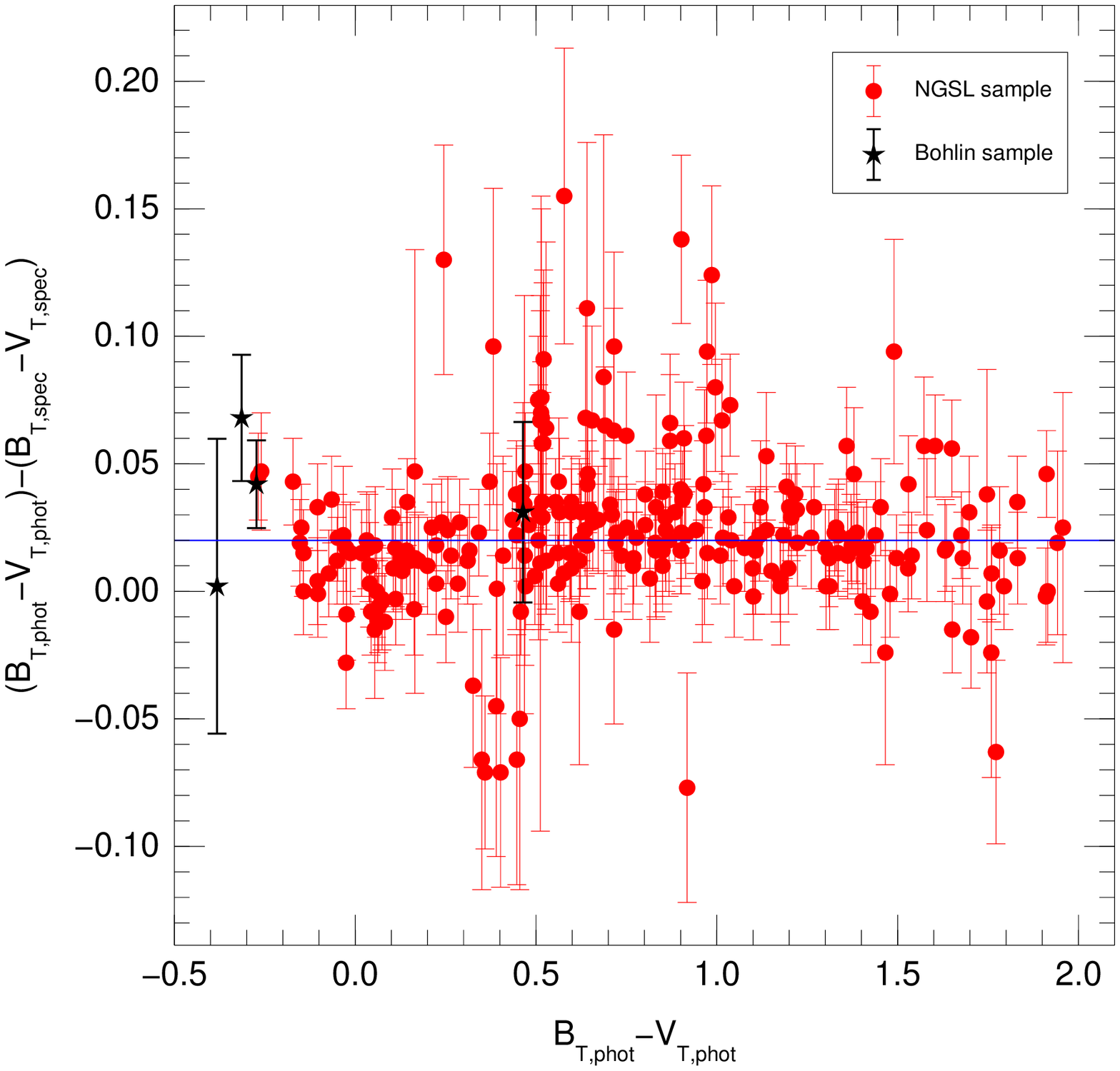}}
\caption{Comparison between photometric and spectrophotometric \btvt\ colors as a function 
of photometric \btvt\ for the two samples. The error bars represent the photometric
uncertainties and the horizontal line marks the proposed ZP$_{B_T-V_T}$.}
\label{btvtplot1}
\end{figure}

\begin{figure}
\centerline{\includegraphics*[width=0.47\linewidth]{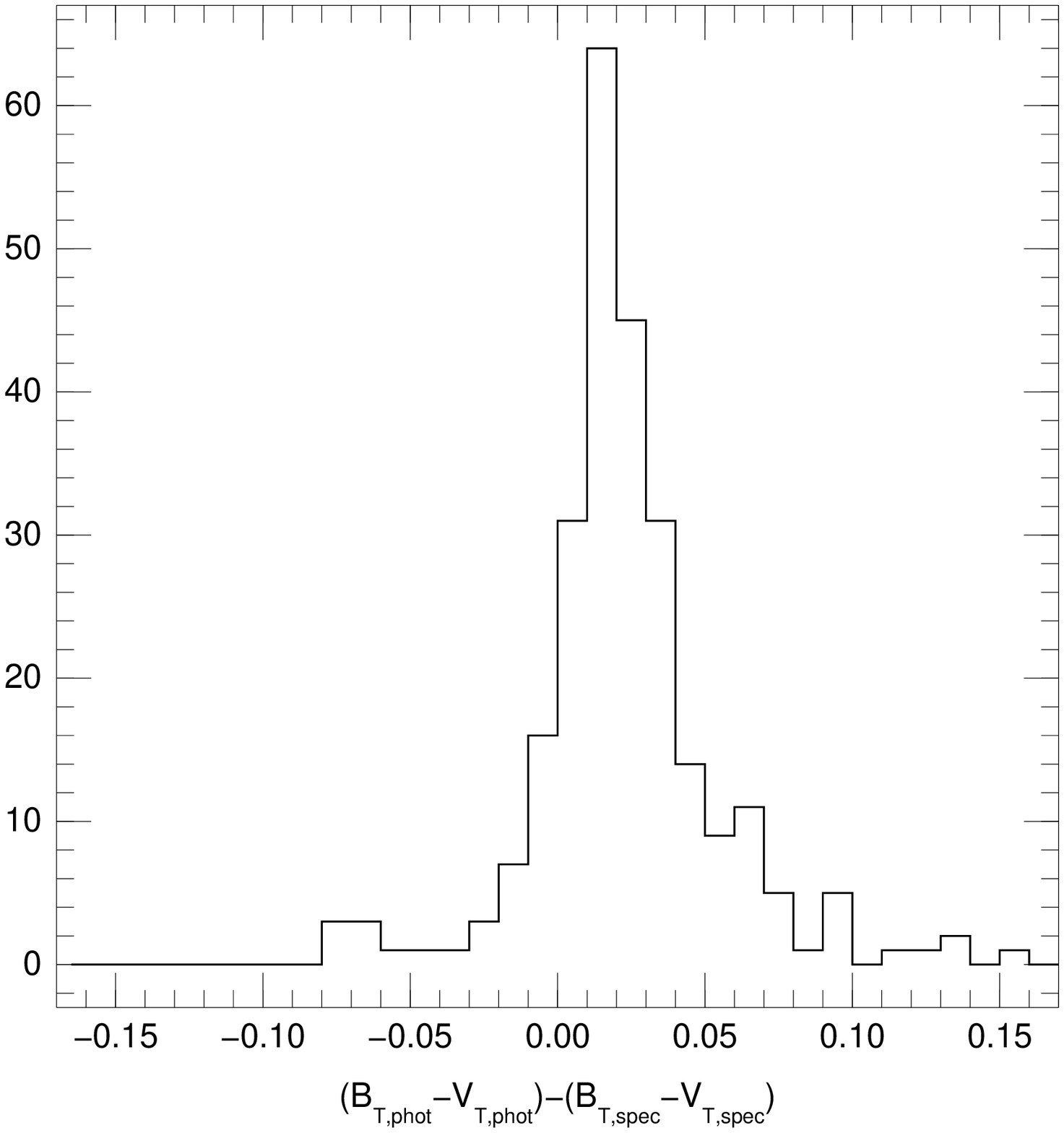}
            \includegraphics*[width=0.47\linewidth]{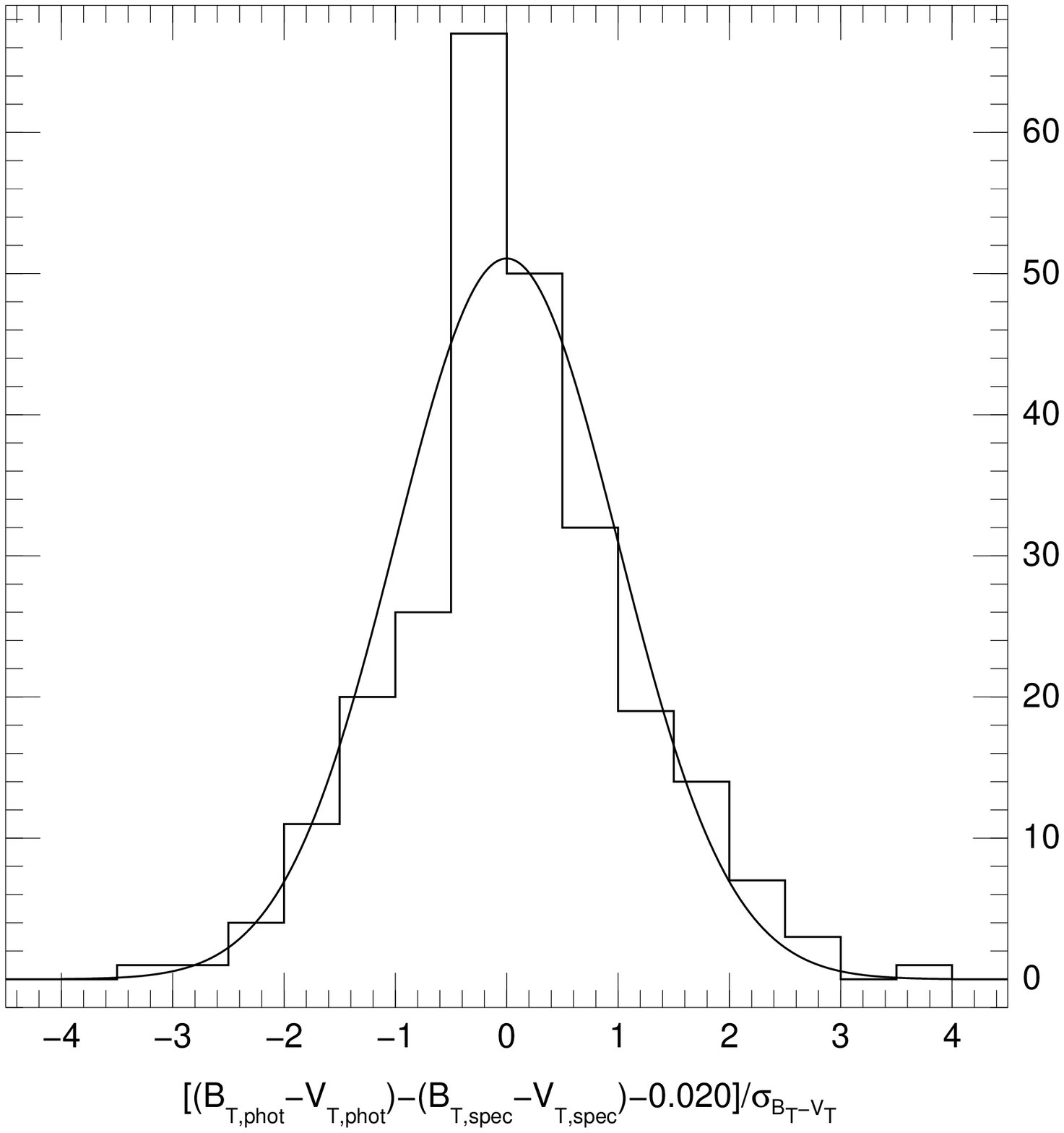}}
\caption{Histograms for the comparison between photometric and spectrophotometric \btvt\ colors
for the NGSL sample. (left) Regular histogram. (right) Histogram for the data shifted by the 
proposed ZP$_{B_T-V_T}$ and normalized by the individual uncertainties. A Gaussian distribution
with $\mu=0$ and $\sigma=1$ is overplotted for comparison.}
\label{btvtplot2}
\end{figure}

	I start by analyzing the relative flux calibration of our data. I show in 
Fig.~\ref{btvtplot1} the difference between the photometric (from the Tycho-2 catalog) and 
spectrophotometric (computed from STIS spectra) values for the \btvt\ of each of the stars in 
our samples as a function of the photometric \btvt. No general trend is observed as a function 
of color and the data are symmetrically distributed around a central value, which I take to be
ZP$_{B_T-V_T}$. I measured ZP$_{B_T-V_T}$ by calculating the weighted mean using 
$1/\sigma^2_{B_T-V_T}$ as weights and found it to be 0.020$\pm$0.001 magnitudes.
I show the histogram for the \btvt\ data in Fig.~\ref{btvtplot2}, both in absolute and
relative (corrected for ZP$_{B_T-V_T}$ and dividing each point by its photometric uncertainty)
terms. The second histogram has a median of $1.2\cdot 10^{-5}$ and a standard deviation of 1.04
and the distribution is very well approximated by a normalized Gaussian. All of the above implies
that an accurate cross-calibration of colors vs. relative fluxes between Tycho-2 photometry and
HST spectrophotometry is possible in principle without having to invoke e.g. modifications in 
the Tycho filter sensitivities or the STIS calibration. Furthermore, given that the normalized
histogram has a standard deviation only slightly larger than 1.0, the largest source of
deviations from the expected value originates in the photometry, not in the spectrophotometry.
Since the mean photometric $\sigma_{B_T-V_T}$ = 0.025 magnitudes, the accuracy of the
spectrophotometrically-derived Tycho-2 colors must be better than 1\%, which agrees with the
published value for the STIS photometric repeatibility \citet{Bohletal01}.

	Only four objects in the Bohlin sample have both \bt\ and \vt\ photometry.
Of course, such a low number is not enough to do
accurate statistics but we can see in Fig.~\ref{btvtplot1} that the results from that sample are
consistent with the derived ZP$_{B_T-V_T}$ and with the absence of a color dependence for the
cross-calibration. One can argue that the two stars with \btvt\ around $-$0.3 (BD +75 325 and 
HD 93521) are slightly above the ZP$_{B_T-V_T}$ line and, hence, may indicate the beginning of a 
small departure from the general rule for very blue objects. However, it can also be a 
fluctuation caused by small-number statistics, which is reinforced by the reversal of the trend 
in the bluest of the stars in the Bohlin sample, BD +28 4211. In any case, even if the effect
turned out to be real, it would be very small and affect only extremely blue stars.

\begin{figure}
\centerline{\includegraphics*[width=1.00\linewidth]{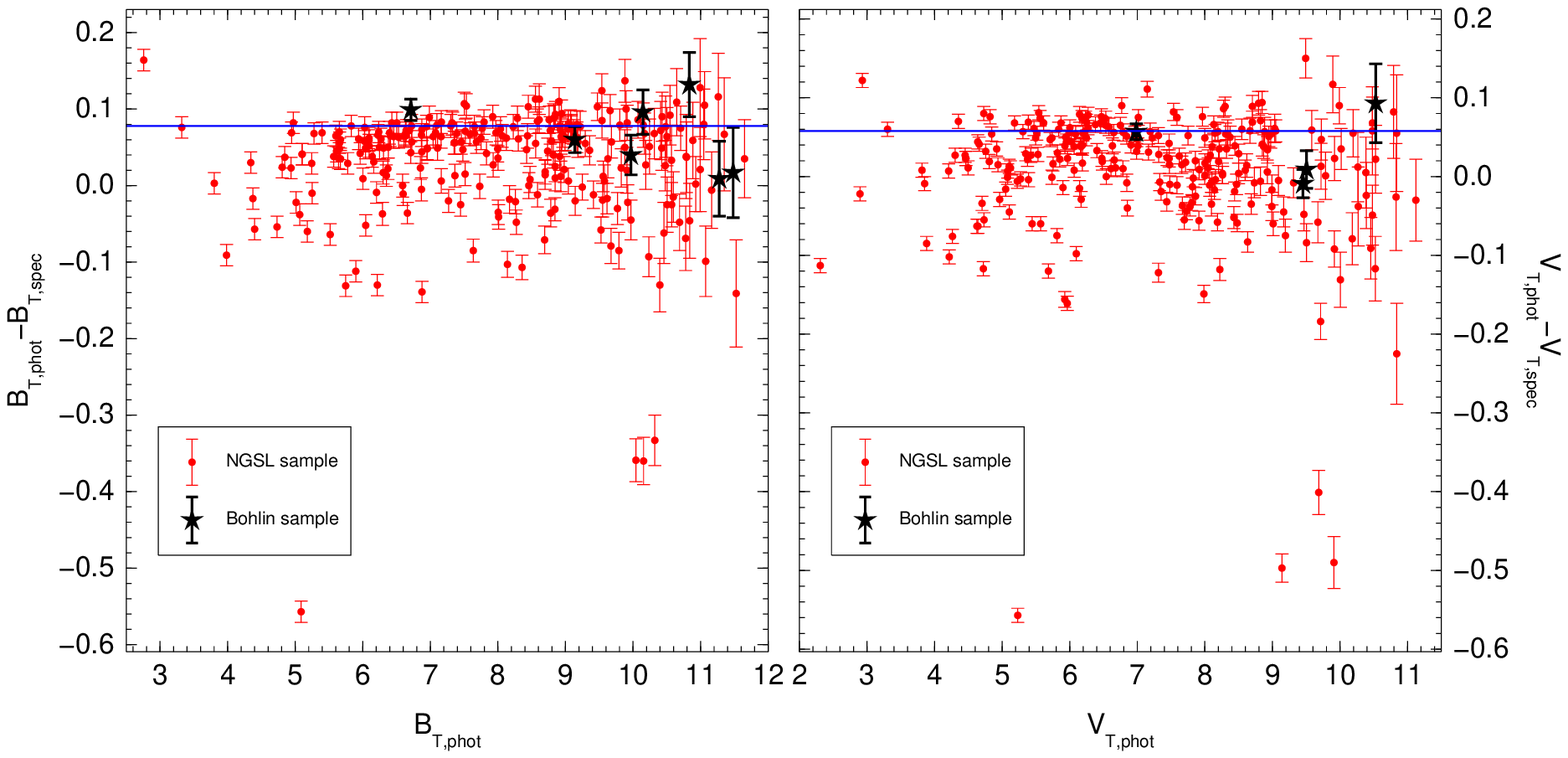}}
\caption{Comparison between photometric and spectrophotometric \bt\ (left) and \vt\ (right)
magnitudes as a function of the photometric values for the two samples. The error bars 
represent the photometric uncertainties and the horizontal line marks the proposed 
ZP$_{B_T}$ (left) and ZP$_{V_T}$ (right). The asymmetric scatter of values below the line is
due to light loss at the slit due to poor centering (see text for details).}
\label{btvtplot3}
\end{figure}

	Next, I deal with the absolute calibration of our data. I show in Fig.~\ref{btvtplot3}
the difference between the photometric and spectrophotometric values for both \bt\ (left panel)
and \vt\ (right panel). Once again, no significant trend is observable as a function of \bt\ or
\vt, thus enabling the cross-calibration of the two systems. However, a difference is readily 
apparent between the relative and absolute plots for the NGSL sample: for the absolute case
the data are also concentrated around a central point but a tail 
towards the negative in the distributions of $B_{T,\rm phot}-B_{T,\rm spec}$ and 
$V_{T,\rm phot}-V_{T,\rm spec}$ is clearly present. Such a behavior is the expected one if some
of the stars are not well centered on the 52$\times$0.2 slit, as previously mentioned. Under
such circumstances it is not possible to derive a value for either ZP$_{B_T}$ or ZP$_{V_T}$
with a precision of the order of 0.01 magnitudes or better. Therefore, for that task I decided 
to use instead the Bohlin sample, which was obtained using the wider 52$\times$2 slit. Given
the high precision achieved for ZP$_{B_T-V_T}$, it is a better strategy to measure just one of
the two absolute ZP values and then use the color zero point to obtain the other one.
Considering that I have more stars with \bt\ data than with \vt\ data (seven vs. four), I
choose the first one for the absolute flux calibration. A weighted mean for those seven stars
in the Bohlin sample yields ZP$_{B_T}$ = 0.078$\pm$0.009 magnitudes. Applying ZP$_{B_T-V_T}$, 
we then obtain ZP$_{V_T}$ = 0.058$\pm$0.009 magnitudes. As it can be seen in
Fig.~\ref{btvtplot3}, those values are compatible with the observed distributions in
$B_{T,\rm phot}-B_{T,\rm spec}$ and $V_{T,\rm phot}-V_{T,\rm spec}$ for the NGSL
sample if one discards the negative-values tail. 

	I have obtained a precise cross-calibration between Tycho-2 photometry and HST
spectrophotometry which, combined with the independent absolute flux calibrations of both
instruments, should allow for accurate comparisons between Tycho-2 photometry and synthetic
photometry generated from SED models. Furthermore, the use of the zero points presented here
should also allow combinations of Tycho-2 photometry with that of other surveys with precise
zero points, such as 2MASS \citep{Coheetal03}.

\acknowledgments

I would like to thank Rodolfo Barb\'a for fruitful conversations on this topic and Ralph Bohlin,
Santiago Arribas, and an anonymous referee for useful comments on the manuscript.

\bibliographystyle{apj}
\bibliography{general}

\end{document}